\documentclass[aps,prx,letter,preprintnumbers,superscriptaddress,nofootinbib,longbibliography,floatfix]{revtex4-2}
\pdfoutput=1
\usepackage{rotating}
\usepackage{array}
\usepackage{amsmath}
\usepackage[normalem]{ulem}
\usepackage{slashed}
\usepackage{booktabs}
\usepackage[pdftex,table]{xcolor}
\usepackage{units}
\usepackage{xfrac}
\usepackage{mathtools}
\usepackage{empheq}
\usepackage[]{units}
\usepackage{multirow}
\usepackage{amssymb}
\usepackage{url}
\usepackage{comment}
\usepackage{physics}
\usepackage{color,soul}
\usepackage{bbm}
\usepackage{adjustbox}
\usepackage[T1]{fontenc}
\usepackage{xspace}

\usepackage{graphicx,caption,subcaption,tikz}
\usetikzlibrary{backgrounds}

\usetikzlibrary{calc}
\usetikzlibrary{positioning}
\DeclareCaptionFormat{overlay}{\gdef\capoverlay{#1#2#3\par}}
\DeclareCaptionStyle{overlay}{format=overlay}
\DeclareCaptionFormat{suboverlay}{\gdef\subcapoverlay{(\thesubfigure) #3\par}}
\DeclareCaptionStyle{suboverlay}{format=suboverlay}

\usepackage{hyperref}
\hypersetup{
  colorlinks=true,
  citecolor=blue,
  linkcolor=blue,
  urlcolor=blue
}

\newcommand{\effval}[1]{%
    \makebox[5.5cm][l]{$\displaystyle #1$}%
}

\begin{document}

\title{Differentiable Parametric Simulation and Reconstruction Models in Parnassus}

\author{Abdelrahman Elabd}
\email{aelabd2@uw.edu}
\affiliation{Department of Physics, University of Washington, Seattle, USA}

\author{Eilam Gross}
\email{eilam.work@gmail.com}
\affiliation{Weizmann Institute of Science, Rehovot, Israel}

\author{Dmitrii Kobylianskii}
\email{dmitry.kobylyansky@weizmann.ac.il}
\affiliation{Weizmann Institute of Science, Rehovot, Israel}

\author{Runze Li}
\email{runze.li@yale.edu}
\affiliation{Department of Physics, Yale University, New Haven, CT 06511, USA}

\author{Benjamin Nachman}
\email{nachman@stanford.edu}
\affiliation{Department of Particle Physics and Astrophysics, Stanford University, Stanford, CA 94305, USA}
\affiliation{Fundamental Physics Directorate, SLAC National Accelerator Laboratory, Menlo Park, CA 94025, USA}

\begin{abstract}
Parnassus is a framework for fast detector simulation and reconstruction, directly mapping truth-level particles onto reconstructed objects. Such models can be built from deep generative networks trained on paired samples, which are fit automatically to a target detector, or from parametric prescriptions of the kind used by Delphes, which are constructed by hand. We remove this asymmetry by making the parametric models fully differentiable so that their parameters can be fit to a target sample by gradient descent. We demonstrate closure by fitting a parametric model to samples from a known configuration of itself, recovering the generating parameters and characterizing the degeneracies among them, and we present a first fit to CMS full simulation. The resulting models are interpretable, inexpensive, and run in the standard Parnassus pipeline which is fully Python based and GPU enabled.
\end{abstract}

\maketitle

\section{Introduction}

Simulations are essential tools in particle physics for connecting theoretical frameworks with experimental data.  Reconstruction reduces the dimensionality of the experimental data to approximate the set of particles impinging on the detector.  The computational cost of simulation and reconstruction are increasingly prohibitive for the most accurate approaches and so fast simulation surrogate models are widely deployed both within and outside of experimental collaborations.  Two types of tools have emerged in this space. Parametric approaches such as Delphes~\cite{deFavereau:2013fsa} replace detailed material modeling and combinatorial reconstruction with analytic smearing functions, encoded in a manually-constructed detector cards. Machine-learned surrogates instead train deep generative models to replace all or parts of the full chain~\cite{Paganini:2017hrr,Paganini:2017dwg,Krause:2024avx,Hashemi:2023rgo,Butter:2022rso,Adelmann:2022ozp,Feickert:2021ajf,Krause:2026ayh}, trading interpretability for fidelity.

Parnassus \cite{Kobylianskii:2024sup,Dreyer:2024bhs,Dreyer:2025zhp, Elabd:2026qfw,Lo:2026use} provides a common interface for simulation and reconstruction models that transform point clouds of truth particles into point clouds of reconstructed objects.  This purely python-based, GPU-enabled framework has two paths: Parnassus-Flow (Parnassus-F) provides conditional flow matching (CFM) models~\cite{lipman2023flowmatchinggenerativemodeling} which are trained on full Geant4-based~\cite{GEANT4:2002zbu} simulation and reconstruction datasets, while Parnassus-Parametrized (Parnassus-P) provides PyTorch\footnote{Previously referred to as TorchDelphes~\cite{Elabd:2026qfw} as a functionally identical PyTorch implementation of Delphes v3.5.1.} implementations of the Delphes detector emulation software~\cite{Elabd:2026qfw}. Downstream processing such as jet clustering, isolation, and analysis-level object definitions is shared across both paths.  The two paths are therefore interchangeable at the level of the pipeline, but they are not on equal footing in one important respect: the neural models are fit, automatically and to a specific target, whereas the parametric models are still assembled by hand, with card parameters (e.g. detector momentum resolution and tracking-efficiency) chosen from performance plots, prior experience, and iteration. Every new detector, every new running condition, and every re-optimization of reconstruction requires that human effort to be repeated. In this work, we close the gap by implementing differentiability for the parametric models and optimizing them against data via gradient descent. 

This asymmetry can be solved if the parametric model is implemented so that its outputs are differentiable with respect to its parameters.  This \textit{differentiable programming} has been explored in a number of areas across particle physics~\cite{Adelmann:2022ozp,MODE:2022znx,MODE:2023ekf}.  For example, there have been a number of proposals for building differentiable physics simulators for gradient-based calibration, which is similar to the goal of Parnassus.  Such studies include liquid argon time projection chambers \cite{Gasiorowski:2023tqf}, optical neutrino detectors \cite{Alterkait:2026ocv}, and calorimeters \cite{aehle2024optimizationusingpathwisealgorithmic,aehle2024efficientforwardmodealgorithmicderivatives}. While these proposals focus on making the microphysical Monte Carlo processes of particular instruments differentiable, their components provide a foundation for the approach studied in this paper.  Such approaches complement another related subject: Monte Carlo generator tuning.  The most well-known tool in this space is Professor~\cite{Buckley:2009bj} and there are also many other approaches that use various forms of surrogate models.

In this paper, we close the asymmetry within Parnassus by making a \textit{differentiable Delphes}, i.e., Parnassus-P is differentiable with respect to its card parameters and we fit those parameters by gradient descent. The output is not a new simulation program but a new card, which runs in the standard Parnassus pipeline like CFM model or a hand-written Delphes card.  The remainder of the paper is organized as follows.   Section~\ref{sec:methods} introduces the technical details of the old (static) and new (differentiable) Parnassus-P framework.  Next, Sec. \ref{sec:pseudodata_closure} validates the new setup through a closure test. Pseudodata are generated from a known, ``target'' parameter configuration of Parnassus-P, then those parameters are perturbed from their target values, and finally the target values are recovered via gradient descent against the generated pseudodata.  Then, we apply Parnassus-P to the CMS detector in Sec.~\ref{sec:opendata}, showing that the resulting model achieves better fidelity, with respect to CMS full simulation, than the default hand-tuned CMS configuration of Delphes.  Finally, we summarize our results and reflect on future directions in Sec.~\ref{sec:conclusions}.

%%%%%

%Parnassus-P was introduced in Ref. \cite{Parnassus-software} as a functionally identical, but not necessarily differentiable, PyTorch implementation of Delphes v3.5.1. This version remains selectable through a user argument. In this work, we implement a differentiable version of Parnassus-P, which diverges slightly from the functionally exact replica of Delphes. Section \ref{sec:methods} provides details on the implementation and optimization of this new differentiable version. 

%We validate the differentiability through closure tests: pseudodata is generated from a known, "target" parameter configuration of TorchDelphes, then those parameters are perturbed from their target values, and finally the target values are recovered via gradient descent against the generated pseudodata. Results from the closure tests are presented in Section \ref{sec:pseudodata_closure}.

%With differentiability implemented and validated, we then train Parnassus-P against full-simulation open data from the CMS \cite{CMS?} detector, and show that the resulting model achieves better fidelity, with respect to CMS full simulation, than the default hand-tuned CMS configuration of Delphes. These results are shown in Section \ref{sec:full_sim}. 

%Finally, we summarize our results and reflect on future directions in Section \ref{sec:conclusion}.

\section{Methods} \label{sec:methods}

\subsection{Differentiability} \label{sec:differentiability}

Delphes does not immediately lend itself to the easy transmission (backpropagation) of meaningful gradient information about the parameters we would like to tune. Therefore, Parnassus-P is designed to be statistically identical to Delphes in a forward (inference) pass, but much more gradient-friendly during a backward (training) pass. This is done primarily through three approaches:
\begin{enumerate}
    \item Endowing the relevant card parameters with automatic gradients by promoting them to instances of PyTorch's \texttt{nn.Parameter} class, and bounding them in gradient-friendly domains. 
    \item Using the reparametrization trick to enable explicit differentiation with respect to smearing/noise parameters (e.g. the scale and width of the log-normal distribution used to smear electron momenta). 
    \item Constructing object-count terms to enable explicit differentiation with respect to both efficiency parameters (e.g. the survival efficiency of charged hadron tracks) and threshold-selection parameters (e.g. minimum significance of a reconstructed calorimeter tower).
\end{enumerate}

The rest of this section elaborates on these approaches.

\subsubsection{Parametrization}\label{sec:parametrization}

Here, we walk through the typical sequence of Delphes modules used to create reconstructed particle candidates, which are also known as Particle Flow Objects (PFOs) based on the Particle Flow reconstruction algorithm~\cite{CMS:2017yfk, ATLAS:2017ghe}. For each module, we list any of its parameters which Parnassus-P implements as auto-differentiable \texttt{nn.Parameter} objects, and elaborate on any functional differences between the module's differentiable and non-differentiable (i.e. Delphes-exact) implementations.

\begin{enumerate}
    \item \texttt{ParticlePropagator} 
    \begin{itemize}
        \item This module takes truth-level particles (generated by event generators such as Pythia~8 ~\cite{sjostrandBriefIntroductionPYTHIA2008, bierlichComprehensiveGuidePhysics2022}) and propagates them through free space and the detector, according to their charge and momentum and the magnetic field inside the detector. \texttt{ParticlePropagator} is not parametrizable in this work, as the detector geometry and magnetic field are treated as constants rather than tunable parameters. 
        
    \end{itemize}
    \item \texttt{\{ElectronTracking/MuonTracking/ChargedHadronTracking\}Efficiency}
    \begin{itemize}
        \item The \texttt{Efficiency} module is generally used to emulate any type of particle loss. A typical Delphes card configures different \texttt{Efficiency} modules to emulate the efficiency with which the tracking subdetector reconstructs tracks from different types of charged particles. For example, the \texttt{ElectronTrackingEfficiency} module takes propagated electrons coming from the \texttt{ParticlePropagator} module and randomly masks each one with a kinematics dependent probability following Eq.~\ref{eq:electron_eff}, emulating the non-ideal efficiency of actual electron tracking. The \texttt{MuonTrackingEfficiency} and \texttt{ChargedHadronTrackingEfficiency} modules do the same thing for muons (Eq.~\ref{eq:muon_eff}) and charged hadrons (Eq.~\ref{eq:chad_eff}), respectively. Each tracking efficiency module introduces several tunable parameters, indicated as $\lambda_0$, $\lambda_1$, or $\texttt{eff\_logits}_*$ tensor elements in the efficiency equations. Parnassus-P therefore provides a total of 18 automatically differentiable parameters for charged tracking efficiency.

        % Instead of long equations we can just draw the scheme.
        % \include{tikz_eff}

        \begin{align}
        \varepsilon_{e}(p_T,\eta)
        &=
        \begin{cases}
            \effval{0}
                & p_T \le 0.1~\mathrm{GeV} \\[2pt]
            \effval{\mathrm{sigmoid}\!\left(\texttt{eff\_logits}_e\texttt{[0]}\right)}
                & |\eta| \le 1.5,\; 0.1 < p_T \le 1~\mathrm{GeV} \\
            \effval{\mathrm{sigmoid}\!\left(\texttt{eff\_logits}_e\texttt{[1]}\right)}
                & |\eta| \le 1.5,\; 1 < p_T \le 100~\mathrm{GeV} \\
            \effval{\mathrm{sigmoid}\!\left(\texttt{eff\_logits}_e\texttt{[2]}\right)}
                & |\eta| \le 1.5,\; p_T > 100~\mathrm{GeV} \\[2pt]
            \effval{\mathrm{sigmoid}\!\left(\texttt{eff\_logits}_e\texttt{[3]}\right)}
                & 1.5 < |\eta| \le 2.5,\; 0.1 < p_T \le 1~\mathrm{GeV} \\
            \effval{\mathrm{sigmoid}\!\left(\texttt{eff\_logits}_e\texttt{[4]}\right)}
                & 1.5 < |\eta| \le 2.5,\; 1 < p_T \le 100~\mathrm{GeV} \\
            \effval{\mathrm{sigmoid}\!\left(\texttt{eff\_logits}_e\texttt{[5]}\right)}
                & 1.5 < |\eta| \le 2.5,\; p_T > 100~\mathrm{GeV} \\[2pt]
            \effval{0}
                & |\eta| > 2.5
        \end{cases}
        \label{eq:electron_eff}
        \\[6pt]
        \varepsilon_{\mu}(p_T,\eta)
        &=
        \begin{cases}
            \effval{0}
                & p_T \le 0.1~\mathrm{GeV} \\[2pt]
            \effval{\mathrm{sigmoid}\!\left(\texttt{eff\_logits}_\mu\texttt{[0]}\right)}
                & |\eta| \le 1.5,\; 0.1 < p_T \le 1~\mathrm{GeV} \\
            \effval{\mathrm{sigmoid}\!\left(\texttt{eff\_logits}_\mu\texttt{[1]}\right)}
                & |\eta| \le 1.5,\; 1 < p_T \le 10^{3}~\mathrm{GeV} \\
            \effval{\mathrm{sigmoid}\!\left(\texttt{eff\_logits}_\mu\texttt{[2]}\right)
                    e^{\,0.5-\lambda_0 p_T}}
                & |\eta| \le 1.5,\; p_T > 10^{3}~\mathrm{GeV} \\[2pt]
            \effval{\mathrm{sigmoid}\!\left(\texttt{eff\_logits}_\mu\texttt{[3]}\right)}
                & 1.5 < |\eta| \le 2.5,\; 0.1 < p_T \le 1~\mathrm{GeV} \\
            \effval{\mathrm{sigmoid}\!\left(\texttt{eff\_logits}_\mu\texttt{[4]}\right)}
                & 1.5 < |\eta| \le 2.5,\; 1 < p_T \le 10^{3}~\mathrm{GeV} \\
            \effval{\mathrm{sigmoid}\!\left(\texttt{eff\_logits}_\mu\texttt{[5]}\right)
                    e^{\,0.5-\lambda_1 p_T}}
                & 1.5 < |\eta| \le 2.5,\; p_T > 10^{3}~\mathrm{GeV} \\[2pt]
            \effval{0}
                & |\eta| > 2.5
        \end{cases}
        \label{eq:muon_eff}
        \\[6pt]
        \varepsilon_{h^\pm}(p_T,\eta)
        &=
        \begin{cases}
            \effval{0}
                & p_T \le 0.1~\mathrm{GeV} \\[2pt]
            \effval{\mathrm{sigmoid}\!\left(\texttt{eff\_logits}_{h^\pm}\texttt{[0]}\right)}
                & |\eta| \le 1.5,\; 0.1 < p_T \le 1~\mathrm{GeV} \\
            \effval{\mathrm{sigmoid}\!\left(\texttt{eff\_logits}_{h^\pm}\texttt{[1]}\right)}
                & |\eta| \le 1.5,\; p_T > 1~\mathrm{GeV} \\[2pt]
            \effval{\mathrm{sigmoid}\!\left(\texttt{eff\_logits}_{h^\pm}\texttt{[2]}\right)}
                & 1.5 < |\eta| \le 2.5,\; 0.1 < p_T \le 1~\mathrm{GeV} \\
            \effval{\mathrm{sigmoid}\!\left(\texttt{eff\_logits}_{h^\pm}\texttt{[3]}\right)}
                & 1.5 < |\eta| \le 2.5,\; p_T > 1~\mathrm{GeV} \\[2pt]
            \effval{0}
                & |\eta| > 2.5
        \end{cases}
        \label{eq:chad_eff}
    \end{align}

    \end{itemize}
    \item \texttt{\{Electron/Muon/ChargedHadron\}MomentumSmearing}
    \begin{itemize}
        \item The \texttt{MomentumSmearing} module is used to emulate the finite momentum resolution of charged particle tracking. Tracking detectors do not yield perfect track information for each successfully tracked particle, but rather a discrete set of finitely granular locations - or "hits" - along that particle's trajectory, which are used to generate a best-fit estimate of the true trajectory. Matching hits and particles is also a source of resolution. The inferred particle (transverse) momenta from these trajectory estimates therefore also have some inherent resolution. 
        
        In Delphes, the \texttt{MomentumSmearing} module emulates this resolution via parametrized log-normal smearing of each particle's true momentum, according to Eq. \ref{eq:mom_smear_default}.
        
        \begin{align}
            p_T^{\mathrm{smeared}} &\sim \mathrm{LogNormal}\!\bigg(\mu = p_T,\; \sigma = \sigma_{p_T}(p_T, \eta)\bigg), \label{eq:mom_smear_default}\\
            &\;\;\;\; \text{where } \nonumber \\ 
            &\;\;\;\;\;\;\;\; \frac{\sigma_{p_T}}{p_T}(p_T, \eta) =
            \begin{cases}
                \sqrt{a_0^2 + b_0^2\, p_T^2} & |\eta| \le 0.5,\; p_T > 0.1~\mathrm{GeV} \\[2pt]
                \sqrt{a_1^2 + b_1^2\, p_T^2} & 0.5 < |\eta| \le 1.5,\; p_T > 0.1~\mathrm{GeV} \\[2pt]
                \sqrt{a_2^2 + b_2^2\, p_T^2} & 1.5 < |\eta| \le 2.5,\; p_T > 0.1~\mathrm{GeV} \\[2pt]
                0 & \text{otherwise}
            \end{cases}
        \end{align}

        The non-differentiable version of Parnassus-P implements this exactly, but the differentiable \texttt{MomentumSmearing} module additionally scales the mean of the smearing distribution by up to $\pm30\%$ in order to model bias effects, and scales the standard deviation by the same amount to avoid distortion. The log-normal momentum smearing of differentiable Parnassus-P therefore follows Eq. \ref{eq:mom_smear}.

        % Can use scheme
        % \include{tikz_res}       
        
        \begin{align}
            p_T^{\mathrm{reco}} &\sim \mathrm{LogNormal}\!\bigg(\mu = s_e(\eta) * p_T,\; \sigma = s_e(\eta) * \sigma_{p_T}(p_T, \eta)\bigg), \label{eq:mom_smear}\\
            &\;\;\;\; \text{where } \nonumber \\ 
            &\;\;\;\;\;\;\;\; \frac{\sigma_{p_T}}{p_T}(p_T, \eta) =
            \begin{cases}
                \sqrt{a_0^2 + b_0^2\, p_T^2} & |\eta| \le 0.5,\; p_T > 0.1~\mathrm{GeV} \\[2pt]
                \sqrt{a_1^2 + b_1^2\, p_T^2} & 0.5 < |\eta| \le 1.5,\; p_T > 0.1~\mathrm{GeV} \\[2pt]
                \sqrt{a_2^2 + b_2^2\, p_T^2} & 1.5 < |\eta| \le 2.5,\; p_T > 0.1~\mathrm{GeV} \\[2pt]
                0 & \text{otherwise}
            \end{cases}  \label{eq:mom_scale} \\ 
            &\;\;\;\; \text{and } \nonumber \\ 
            &\;\;\;\;\;\;\;\; s_{e}(\eta) = \begin{cases}
                1 + 0.3\tanh\!\left(s_0\right) & |\eta| \le 0.5 \\[2pt]
                1 + 0.3\tanh\!\left(s_1\right) & 0.5 < |\eta| \le 1.5 \\[2pt]
                1 + 0.3\tanh\!\left(s_2\right) & 1.5 < |\eta| \le 2.5 \\[2pt]
                1 & \text{otherwise}
            \end{cases}
        \end{align}

        As with the \texttt{TrackingEfficiency} modules, a typical Delphes card has uniquely parametrized configurations for electrons, muons, and charged hadrons. In Parnassus-P, each is parametrized by 9 values ($\{a_0, a_1, a_2, b_0, b_1, b_2, s_0, s_1, s_2\})$, resulting in a total of 27 automatically differentiable momentum-smearing parameters. 
            
    \end{itemize}
    \item \texttt{\{Track\}Merger}
    \begin{itemize}
        \item The \texttt{Merger} module simply combines different types of objects within an event into a single collection. The \texttt{TrackMerger} configuration of this module combines the post-\texttt{Efficiency} and post-\texttt{MomentumSmearing} tracks for electrons, muons, and charged hadrons into a single set of charged tracks.
    \end{itemize}
    \item \texttt{SimpleCalorimeter}
    \begin{itemize}
        \item The \texttt{SimpleCalorimeter} module has two uniquely configured instantiations: \texttt{ECal}, which emulates the electromagnetic calorimeter, and \texttt{HCal}, which emulates the hadronic calorimeter. This module emulates the deposition of energy by particles in the calorimeters, and the binning of unique $(\eta, \phi)$ tower hits into disjoint hit clusters. It also handles part of the reconstruction step, inferring the existence of neutral particles (i.e. photons and neutral hadrons) from the combination of track and calorimeter information - mimicking the ParticleFlow reconstruction algorithm~\cite{CMS:2017yfk, ATLAS:2017ghe}.

        % The fraction of a particle's energy that is deposited in the electromagnetic calorimeter is determined by its species. In Delphes, this is just a constant for each particle species; Parnassus-P keeps these constants for all but 5 species, which are instead parametrized according to Eq. \ref{eq:hadron_frac}. The $K^0_S$, $\gamma$, $\pi^0$, and $K^0_L$ fractions are confined to narrower, more physical windows than the nominal $(0,1)$. 

        % \begin{equation}
        %     f_{\mathrm{ECal}}(\mathrm{PID}) =
        %         \begin{cases}
        %         \mathrm{sigmoid}\!\left(\ell_{h^\pm}\right) & h^\pm \; (\text{charged hadron}) \\[3pt]
        %         0.1 + 0.4 *\mathrm{sigmoid}\!\left(\ell_{K^0_S}\right) & K^0_S \; (\text{K-short strange meson}) \\[3pt]
        %         \mathrm{sigmoid}\!\left(\ell_{\Lambda}\right) & \Lambda \; (\text{Lambda baryon}) \\[3pt]
        %         0.8 + 0.2 * \mathrm{sigmoid}\!\left(\ell_{\gamma}\right) & \gamma,\, \pi^0 \; (\text{photon, neutral pion})\\[3pt]
        %         0.4 * \mathrm{sigmoid}\!\left(\ell_{K^0_L}\right) & K^0_L \; (\text{K-long strange meson}) \\[3pt]
        %         \text{default constant} & \text{otherwise} 
        %         \end{cases}
        %     \label{eq:hadron_frac}
        % \end{equation}

        The fraction of a particle's energy that is deposited in the electromagnetic calorimeter is determined by its species. In Delphes, this is a constant for each particle species.
        Certain species, such as muons, neutrinos, and BSM neutralinos, deposit zero energy in either calorimeter. For all others, the energy deposited in the hadronic calorimeter is, by construction, all of the energy that is not deposited in the electromagnetic calorimeter, i.e. Eq. \ref{eq:hcal_frac}.

        \begin{equation}
            f_{\mathrm{HCal}}(\mathrm{PID}) =
                \begin{cases}
                0 & \mu, \nu, \tilde{N} \; (\text{muon, neutron, neutralino}) \\[3pt]
                1 - f_{\mathrm{ECal}}(\mathrm{PID}) & \text{otherwise}
                \end{cases}
            \label{eq:hcal_frac}
        \end{equation}

         The energy deposited in each tower is then smeared according to Eq.~\ref{eq:calo_smear}, in direct analogy to the momentum smearing of Eq.~\ref{eq:mom_smear}. As with \texttt{MomentumSmearing}, Parnassus-P introduces a per-region multiplicative energy scale (Eq.~\ref{eq:ecal_scale}, \ref{eq:hcal_scale}) that is absent from Delphes and fixed to unity in the non-differentiable configuration. Note that, unlike the momentum scale, the calorimeter scale is applied to the tower energy \textit{before} the resolution is evaluated, so it does not leave the relative resolution invariant.

        % Can replace by figure
        % \include{tikz_cal}
        
        \begin{align}
            E_X^{\mathrm{smeared}} & \sim \mathrm{LogNormal}\!\bigg(
                \mu = s_X(\eta) * E_X, \;
                \sigma =  \sigma_X\big( s_X(\eta) * E_X, \eta\big)\bigg),
                \qquad X \in \{\mathrm{ECal},\, \mathrm{HCal}\}
            \label{eq:calo_smear} \\
            &\text{where } \nonumber \\
            & \;\;\;\; \sigma_\text{ECal} (E, \eta) =
            \begin{cases}
            \left(a_b + b_b\, \eta^2\right)\sqrt{(c_E^\text{ECal})^2 E^2 + (c_S^\text{ECal})^2 E + (c_N^\text{ECal})^2} & |\eta| \le 1.5 \\[3pt]
            \left(a_e + b_e\, (|\eta| - 2)^2\right)\sqrt{(c_E^\text{ECal})^2 E^2 + (c_S^\text{ECal})^2 E + (c_N^\text{ECal})^2} & 1.5 < |\eta| \le 2.5 \\[3pt]
            \sqrt{(f_E^\text{ECal})^2 E^2 + (f_S^\text{ECal})^2 E} & 2.5 < |\eta| \le 5.0
            \end{cases}
            \label{eq:ecal_res} \\
            & \;\;\;\; \sigma_\text{HCal}(E, \eta) =
            \begin{cases}
            \sqrt{(c_E^\text{HCal})^2 E^2 + (c_S^\text{HCal})^2 E} & |\eta| \le 3.0 \\[3pt]
            \sqrt{(f_E^\text{HCal})^2 E^2 + (f_S^\text{HCal})^2 E} & 3.0 < |\eta| \le 5.0
            \end{cases}
            \label{eq:hcal_res} \\
            & \;\;\;\; s_\text{ECal}(\eta) =
            \begin{cases}
            1 + 0.3\tanh\!\left(t_0\right) & |\eta| \le 1.5 \\[3pt]
            1 + 0.3\tanh\!\left(t_1\right) & 1.5 < |\eta| \le 2.5 \\[3pt]
            1 + 0.3\tanh\!\left(t_2\right) & |\eta| > 2.5
            \end{cases} \label{eq:ecal_scale} \\
            & \;\;\;\; s_{\mathrm{HCal}}(\eta) =
            \begin{cases}
            1 + 0.3\tanh\!\left(u_0\right) & |\eta| \le 3.0 \\[3pt]
            1 + 0.3\tanh\!\left(u_1\right) & |\eta| > 3.0
            \end{cases} \label{eq:hcal_scale} 
        \end{align}

        In can be found that, in Eq.~\ref{eq:ecal_res}, when parameters ($a_b$, $b_b$, $a_e$, $b_e$) are scaled by $\kappa$ and parameters ($c^{ECal}_E$, $c^{ECal}_S$, $c^{ECal}_N$) are scaled by $\frac{1}{\kappa}$, the equation stays unchanged. This property raises a degeneracy in those parameters, which is resolved by freezing one of the parameters, $a_b$ in this case, at constant value 1.0.
        As a result, in total \texttt{SimpleCalorimeter} contributes 17 unique tunable parameters: 
        %the 5 $\ell_*$ parameters of Eq.~\ref{eq:hadron_frac}, 
        the 8 $a_*, b_*, c_*^\text{ECal}$, and $f_*^\text{ECal}$ parameters of Eq.~\ref{eq:ecal_res}, the 4 $c_*^\text{HCal}$ and $f_*^\text{HCal}$ parameters of Eq.~\ref{eq:hcal_res}, the 3 $t_*$ parameters of Eq.~\ref{eq:ecal_scale}, and the 2 $u_*$ parameters of Eq.~\ref{eq:hcal_scale}.

    \end{itemize}
    \item \texttt{Calorimeter} 
    \begin{itemize}
        \item This instance of the \texttt{Merger} module is configured to merge the \texttt{ECal} and \texttt{HCal} tower outputs into a single collection.
    \end{itemize}
    \item \texttt{EFlowMerger}
    \begin{itemize}
        \item This instance of the \texttt{Merger} module is configured to merge calorimeter and track outputs into a single collection of reconstructed particles, each classified as an electron, muon, charged hadron, neutral hadron, or photon.
    \end{itemize}
\end{enumerate}

In total, Parnassus-P provides 62 automatically-differentiable parameters: 18 from the charged particle tracking, 27 from the charged particle momentum smearing, and 17 from the electromagnetic and hadronic calorimeters.

\subsubsection{Reparametrization trick}

The smearing of momentum and energy values introduces a difficulty when it comes to gradient propagation. Consider an electron with true transverse momentum $p_T^\text{el}$. To smear this momentum as described in section \ref{sec:parametrization}, the straightforward approach is to draw a random variable $X$ from a log-normal distribution centered at $p_T^\text{el}$ (plus or minus some bias), 
\begin{align}
    p_T^\text{el, smeared} &\sim \text{LogNormal}\!\bigg(\mu^\text{el}(p_T^\text{el}), \sigma^\text{el}(p_T^\text{el}) \bigg), \label{eq:electron_smearing_draw}
\end{align}
where $\mu^\text{el}$ and $\sigma^\text{el}$ have some explicit dependence on $p_T^\text{el}$ and the tunable parameters of interest, as in Eq. \ref{eq:mom_smear}.
$X$, being a direct random variable, has no explicit dependence on $\mu^\text{el}$, nor $\sigma^\text{el}$, nor the tunable parameters, and so it will not contribute any information about them to the gradient. Moreover, the source distribution changes with each draw since $p_T^\text{el}$ itself, and therefore $\mu^\text{el}$ and $\sigma^\text{el}$, also varies. Therefore, even in the limit of many draws we still cannot necessarily expect meaningful gradient information about the parameters in the realization of $p_T^\text{el, smeared}$.

The solution is a common workaround called the \textit{reparametrization trick}. We instead draw a random variable $Y$ from a standard normal distribution:
\begin{equation}
    Y \sim \mathcal{N}(0,1)
\end{equation}

And transform it into $X'$ according to Eq. \ref{eq:reparam_trick}

\begin{equation}
    X' = e^{m+s*Y} \qquad \text{where } m = \ln \mu^\text{el} - \frac{1}{2}\ln \left(1 + \frac{(\sigma^\text{el})^2}{(\mu^\text{el})^2} \right) \text{ and } s = \sqrt{\ln \left(1+ \frac{(\sigma^\text{el})^2}{(\mu^\text{el})^2}\right)}\label{eq:reparam_trick}
\end{equation}

One can check that the resulting $X'$ is distributed exactly according to Eq. \ref{eq:electron_smearing_draw}. Moreover, unlike $X$, $X'$ has explicit dependence on $\mu^\text{el}$ and $\sigma^{\text{el}}$, and therefore the tunable parameters. 

This reparametrization trick is implemented for all log-normal smearing of energy and momentum values. 

\subsubsection{Object counts and efficiencies}
\label{sec:obj_count}

Object-rejection operations raise another difficulty. Consider the \texttt{TrackingEfficiency} modules; in a forward pass, every track $i$ in region $r_i$ is kept or dropped by a Bernoulli draw applied to its momentum with probability $\epsilon_{r_i}$,
\begin{equation}
    m_i \sim \text{Bernoulli}(\epsilon_{r_i}), \qquad p_{T,i} \rightarrow m_i * p_{T,i}, \qquad m_i \in\{0, 1\} \label{eq:decision_mask}
\end{equation}

For a loss function that only compares particle kinematics (see Section \ref{sec:model_training}), there is no explicit dependence on the parameter of interest, $\epsilon_{r_i}$. As with the log-normal smearing discussed in the previous section, the dependence is at best statistical. Moreover, one cannot simply route $\epsilon_{r_i}$ into the gradient by directly multiplying it with the kinematics,
\begin{equation}
    p_{T,i} \rightarrow \epsilon_{r_i} * p_{T,i} \label{eq:efficiency_mask}
\end{equation}

Not only does this physically rescale the surviving tracks, it also inverts the gradient's sign in the low-$p_T$ regions. With the decision-mask multiplication (Eq. \ref{eq:decision_mask}), increasing the efficiency of a low-$p_T$ region would increase the number of surviving low-$p_T$ tracks and decrease the mean $p_T$ of the event; with the efficiency-mask multiplication (Eq. \ref{eq:efficiency_mask}), increasing the efficiency of a low-$p_T$ region would increase the contributed $p_T$ from every low-$p_T$ track and instead \textit{increase} the mean $p_T$ of the event. Therefore, a straight-through gradient of $\epsilon_{r_i}$ through the particle kinematics is not viable. 

Instead, gradients are passed to the efficiency parameters by introducing a loss on the difference between the expected and observed number of objects (in each efficiency region). Each reconstructed track is assigned two labels on the same $(p_T, |\eta|)$ grid: the pre-reconstruction region from which its efficiency was evaluated, bin $r$, and the post-reconstruction region to which it is finally assigned, bin $b$. The two differ when reconstruction moves a track across a bin boundary.
The number of reconstructed tracks in $b$ originating from $r$ can then be defined as $M_{br}$, which is gradient detached. 
The differentiable expected count in reconstructed region $b$ is
\begin{equation}
    \hat N_b = \sum_r \frac{\epsilon_r}{\text{sg}[\epsilon_r]}\, M_{br}, \label{eq:expected_count}
\end{equation}
where $\text{sg}$ is the "stop-gradient" operator ($\texttt{detach}$ in PyTorch), which detaches gradients from the function argument. $\hat N_b$ is numerically equivalent to $\sum_r M_{br}$ but has $\partial \hat N_b / \partial \epsilon_r = M_{br}/\epsilon_r$. 
Since the expectation of $M_{br}$ is proportional to $\epsilon_r$, $M_{br}/\epsilon_r$ is an unbiased estimate of the derivative of the expected count.
The object rate per event in region $b$ is estimated as $\hat n_b = \hat N_b / N_\text{events}$ in each batch, and is compared to the observed rates $n_b^\text{data}$ of reconstructed tracks in the target sample through a normalized $\chi^2$ for each track species (i.e. electrons, muons, and charged hadrons),
\begin{equation}
    \mathcal{L}_\text{count} = \frac{w_\text{count}}{\sum_b n_b^\text{data}} \sum_b \frac{\left(\hat n_b - n_b^\text{data}\right)^2}{n_b^\text{data} + f}, \label{eq:count_loss}
\end{equation}
with $f = 0.05$ a floor for sparsely populated bins and $w_\text{count}$ the weight relative to the other terms of the loss. This is the only gradient path to the efficiency logits.

A similar issue arises in the calorimeters, and a similar fix is implemented. Each tower must pass a threshold on its energy and on the statistical significance of that energy,

\begin{equation}
E > E_\mathrm{min}
\quad\text{and}\quad
E > n_\sigma\, \sigma_E(E, \eta),
\label{eq:tower_cut}
\end{equation}
where $E$ is the smeared tower energy, $\sigma_E(E, \eta)$ is the resolution of Eqs. \ref{eq:ecal_res} and Eq. \ref{eq:hcal_res}, and $E_\mathrm{min}$ and $n_\sigma$ are fixed thresholds (i.e. constants, not tunable parameters). The resolution parameters enter the second condition through the threshold itself; varying them will widen or narrow $\sigma_E$ and thereby add or remove towers from the selection.

A surviving tower may then produce a neutral object, meaning a photon or neutral hadron inferred from calorimeter energy in excess of what the matched tracks account for. This happens if

\begin{equation}
E_\mathrm{neut} > E_\mathrm{min}
\quad\text{and}\quad
E_\mathrm{neut} > n_\sigma \sqrt{\sigma_\mathrm{track}^2 + \sigma_E^2},
\qquad
E_\mathrm{neut} = \max\left(E - E_\mathrm{track},\, 0\right),
\label{eq:neutral_cut}
\end{equation}
where $E_\mathrm{track}$ is the summed energy of the tracks matched to the tower and $\sigma_\mathrm{track}$ is its resolution, obtained by adding the momentum resolutions in quadrature over those tracks. Eqs. \ref{eq:tower_cut} and \ref{eq:neutral_cut} together determine whether a neutral object is produced. In analogy to the tracking efficiencies, we compare the expected and observed counts of these neutral objects (i.e. photons and neutral hadrons) through a loss term in the form of Eq. \ref{eq:count_loss}. Each hard threshold comparison is replaced with a sigmoid that turns on smoothly as the threshold is surpassed, using a transition width of $\tau = 0.05$ times the threshold value. Their product is a per tower probability $g_t$ that a neutral object survives, so the expected count in an $|\eta|$ region $k$ is
\begin{equation}
    \hat N_k = \sum_{t \, \in \, k} g_t. \label{eq:calo_expected_count}
\end{equation}
The two neutral-object count terms supply a gradient for the calorimeter resolution parameters of Eqs. \ref{eq:ecal_res} and \ref{eq:hcal_res}, though this is not necessarily the only gradient those parameters receive, unlike the tracking efficiency parameters.

\subsection{Model Training} \label{sec:model_training}

Parnassus-P takes as input the ($p_\mathrm{T}$, $\eta$, $\phi$, PDG id) of the truth particles in an event, from which the charge and mass used in propagation and energy deposition are derived, and outputs the ($p_\mathrm{T}$, $\eta$, $\phi$, class) of the reconstructed particle-flow objects, where class $\in$ \{charged hadron, electron, muon, neutral hadron, photon\}. Events within a batch are zero-padded to the largest particle multiplicity in the batch, so that the input is a dense tensor of shape (batch size, $N^{\mathrm{pad}}_{\mathrm{particles}}$, 4), where the last dimension holds the four input features listed above. Padded entries are masked and do not contribute to the simulation or the loss.

The loss used to train Parnassus-P contains several terms, summarized in Table~\ref{tab:loss_terms}. The first group consists of Wasserstein distances between the one-dimensional simulated and target distributions of each reconstructed observable, with objects pooled over all events in the batch and evaluated separately for each object class.
In one dimension, the Wasserstein distance has a closed form:
$$W_2^2(\mu,\nu) = \int_0^1 \left[F_\mu^{-1}(u) - F_\nu^{-1}(u)\right]^2 \mathrm{d}u$$
with $F^{-1}$ the quantile functions of the two distributions~\cite{villani2009optimal}, so it can be evaluated exactly from the sorted samples, without any binning or solving the optimal transport problem. 
Energies and momenta enter the loss as $\log E$ and $\log p_T$ rather than in linear units. Because the spectra fall steeply over several orders of magnitude, the Wasserstein distance in linear units is dominated by a few high $p_T$ objects and is insensitive to the bulk of the population.
The $\log E$ and $\log p_T$ terms are further split by $|\eta|$ region. In Delphes, a single parameter group typically contains separate parameters for different $|\eta|$ regions, and splitting the terms gives the parameters governing the sparsely populated endcap and forward regions a gradient of their own, which would otherwise be swamped by the densely populated central region. 
The second group consists of event level terms: the distribution of total transverse energy $\log H_T$, and the response $\ln(m^{\mathrm{reco}}/m^{\mathrm{truth}})$, which is the ratio between the invariant mass of the two leading reconstructed muons, electrons, or charged hadrons and the invariant mass of the corresponding truth pair in the same event. 
The width of the response distribution directly measures the momentum resolution and momentum scale, which are not visible in the single object distributions of a smooth spectrum. 
Finally, because the Wasserstein distance is insensitive to the overall normalization, the relative $\chi^2$ terms are computed between the mean number of reconstructed objects per event in the simulated and target samples for each object class and detector region, as explained in Sec.~\ref{sec:obj_count}. 
These terms provide the gradient for the tracking efficiencies, and they offer an additional constraint on the calorimeter parameters.

In Parnassus-P, each detector parameter is stored as an unconstrained raw value and is mapped to its physical range by a sigmoid function (efficiencies and energy fractions), a softplus function (resolution coefficients), or a bounded $\tanh$ function (momentum and energy scales). 
Rather than being initialized randomly, the parameter values are initialized according to the Delphes default CMS card, offering a more physically meaningful starting point for the model.
Because the natural step size can be different for different parameters, these parameters are divided into three groups: resolution, scale, and efficiency. Each group has its own learning rate.
The three learning rates are optimized with Optuna~\cite{akiba2019optuna} search.
The events are split into training, validation, and test sets in the proportion 70\%/20\%/10\%. 
Parnassus-P is trained with the Adam optimizer~\cite{adam2017}.
Training uses a global batch size of 2048 events. 
The learning rates are halved whenever the validation loss has not improved for four epochs, and the training is stopped after ten epochs without improvement, up to a maximum of 100 epochs.

\begin{table}[t]
\centering
\captionsetup{justification=raggedright,singlelinecheck=false}
\setlength{\tabcolsep}{8pt}
\begin{tabular}{lll}
\toprule
Block & Term & Regions \\
\midrule
\multirow{3}{*}{Objects}
 & $\log E$   & per object class and $|\eta|$ region \\
 & $\log p_T$ & per object class and $|\eta|$ region \\
 & $\eta$     & per object class \\
\midrule
\multirow{4}{*}{Events}
 & $\log H_T$ & per event \\
 & $\ln(m^{\mathrm{reco}}_{\mu^+\mu^-}/m^{\mathrm{truth}}_{\mu^+\mu^-})$ & per truth mass window and $|\eta|$ region\\
 & $\ln(m^{\mathrm{reco}}_{e^+e^-}/m^{\mathrm{truth}}_{e^+e^-})$ & per truth mass window and $|\eta|$ region \\
 & $\ln(m^{\mathrm{reco}}_{h^+h^-}/m^{\mathrm{truth}}_{h^+h^-})$ & per truth mass window and $|\eta|$ region\\
\midrule
\multirow{2}{*}{Counts}
 & $\mathcal{L}^{track}_\text{count}$ & per charged object class and $(p_T, |\eta|)$ region \\
 & $\mathcal{L}^{calo}_\text{count}$  & per neutral object class and $|\eta|$ region \\
\bottomrule
\end{tabular}
\caption{Composition of the loss used to train Parnassus-P. Object-level terms: for each object class $c$, the Wasserstein distance between the simulated and target distributions of $\log E$ and $\log p_T$ is computed separately in each $|\eta|$ region, and that of $\eta$ over the full acceptance. Event-level terms: the Wasserstein distance between the simulated and target distributions of $\log H_T$ per event, and, for the two leading-$p_T$ reconstructed objects of each of the $\mu^\pm$, $e^\pm$, and charged-hadron classes, the Wasserstein distance between the simulated and target distributions of the pair-mass response $\ln(m^{\mathrm{reco}}/m^{\mathrm{truth}})$, where $m^{\mathrm{truth}}$ is the mass of the leading truth pair of the same class in the same event; these terms are evaluated separately in windows of $m^{\mathrm{truth}}$ and $|\eta|$ regions. Count terms: the relative $\chi^2$ between the mean number of reconstructed objects per event in the simulated and target samples, evaluated for each charged class in the $(p_T, |\eta|)$ bins of the tracking-efficiency parametrization and for each neutral class in $|\eta|$ bins. All terms are summed with fixed relative weights.}
\label{tab:loss_terms}
\end{table}

\section{Results}\label{sec:results}

\subsection{Closure Tests}
\label{sec:pseudodata_closure}

To demonstrate the performance of Parnassus-P, we perform a closure test by fitting it to a hidden Delphes card, which has been randomly perturbed from the default CMS card values. The perturbed card is used to generate reconstructed samples, which Parnassus-P is then trained on, and the parameter convergences towards their truth values in the perturbed card is checked.

\begin{figure}[h!]
\centering
  \begin{subfigure}[b]{0.38\textwidth}
    \centering
    \includegraphics[width=0.8\linewidth,page=2]{images/example_reg.pdf}
    \caption{Train and validation loss}\label{fig:muon_param_reg_a}
  \end{subfigure}%               <- the % is essential: 0.5 + 0.5 + a space overflows the line
  \begin{subfigure}[b]{0.38\textwidth}
    \centering
    \includegraphics[width=0.8\linewidth,page=1]{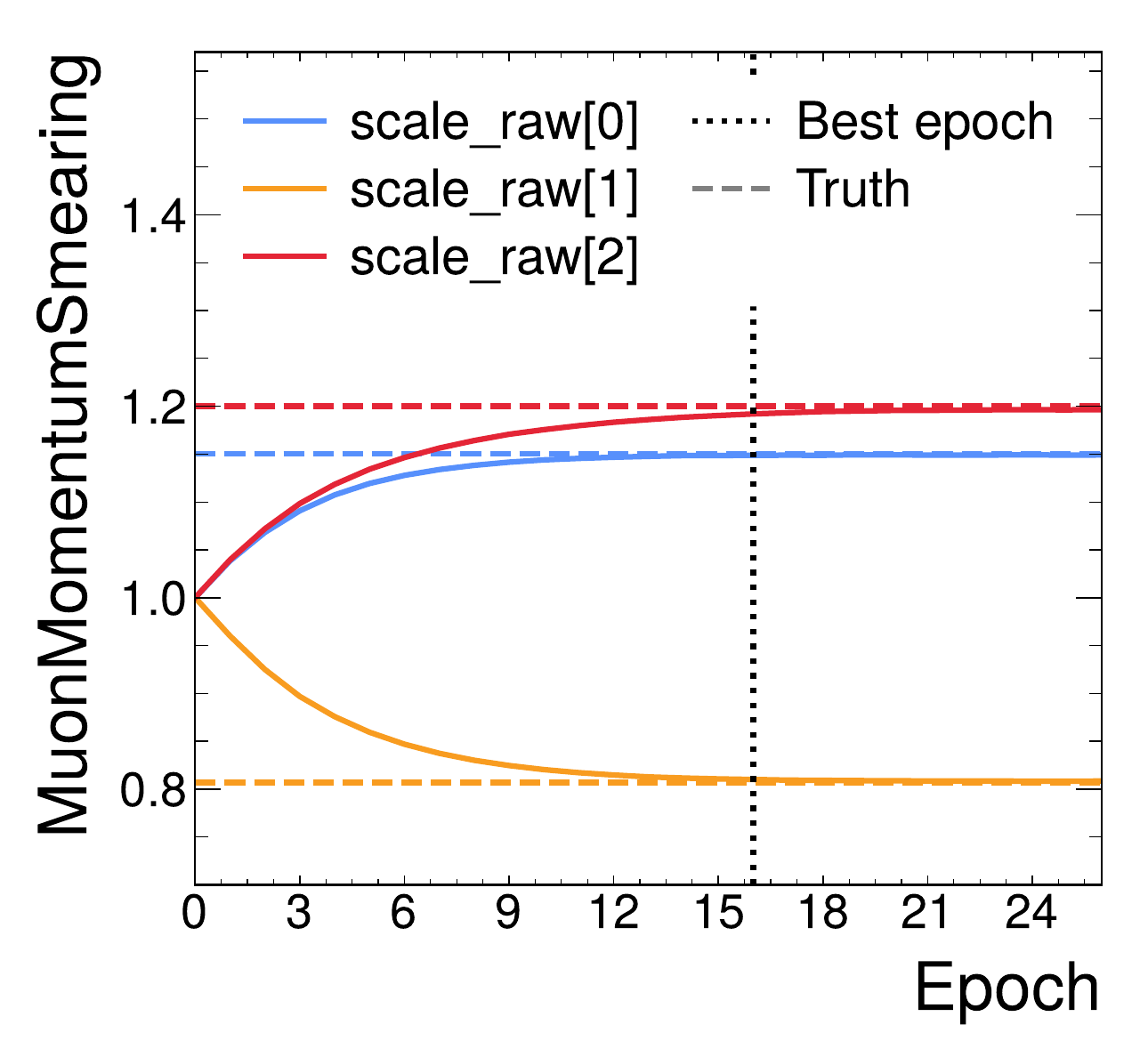}
    \caption{Parameter value}\label{fig:muon_param_reg_b}
  \end{subfigure}
    \caption{Plots of the loss and \texttt{MuonMomentumSmearing.scale\_raw} parameter values during training. The parameters are initialized at their default CMS card values, and trained on 200,000 $Z\rightarrow\mu^+\mu^-$ and $J/\psi\rightarrow\mu^+\mu^-$ events that were generated with the perturbed card configuration. The training early stops at epoch 16, and the parameter values there are taken to be the final result.}
    \label{fig:muon_param_reg}
\end{figure}

As an example, Fig.~\ref{fig:muon_param_reg} shows the training dynamics of Parnassus-P. Initially, the muon tracking efficiency and momentum smearing parameters of Eqs.~\ref{eq:muon_eff} and~\ref{eq:mom_smear} are perturbed. This perturbed configuration is used to generate 200,000 $Z\rightarrow\mu^+\mu^-$ and $J/\psi\rightarrow\mu^+\mu^-$ events, which the model then trains on. The training lasts for 16 epochs, with Fig~\ref{fig:muon_param_reg_a} showing the loss and Fig~\ref{fig:muon_param_reg_b} showing the values of three selected parameters. All three parameters recover their truth values from the hidden Delphes card.

A more comprehensive closure test over all tunable parameter is performed. 
In this setup, all 62 tunable parameters are perturbed away from their default values.
This perturbed card is used to generate 200,000 events for each of the four samples: $Z\rightarrow\mu^+\mu^-$ and $J/\psi\rightarrow\mu^+\mu^-$, $Z\rightarrow e^+e^-$ and $J/\psi\rightarrow e^+e^-$, $K^0_{s} \rightarrow \pi^+ \pi^-$, and QCD dijets with $p_\mathrm{T}$ > 20 GeV. 
These samples are used to sequentially train subsets of the Parnassus-P parameters, as shown in Tab.~\ref{tab:tuning_procedure}.

\begin{table}[h!]
    \centering
    \captionsetup{justification=raggedright,singlelinecheck=false}
    \begin{tabular}{c l l l}
        \toprule
        Step & Calibration sample & Tuned block & Parameters / equations \\
        \midrule
        1 &
        $Z\rightarrow\mu^+\mu^-$,
        $J/\psi\rightarrow\mu^+\mu^-$ &
        Muon tracking &
        $\varepsilon_\mu$ (Eq.~\ref{eq:muon_eff}),
        momentum smearing (Eq.~\ref{eq:mom_smear})
        \\

        2 &
        $K^0_S\rightarrow\pi^+\pi^-$ &
        Charged-hadron tracking &
        $\varepsilon_{h^\pm}$ (Eq.~\ref{eq:chad_eff}),
        momentum smearing (Eq.~\ref{eq:mom_smear})
        \\

        3 &
        Dijet &
        ECal and HCal &
        ECal resolution (Eq.~\ref{eq:ecal_res}),
        HCal resolution (Eq.~\ref{eq:hcal_res})
        \\

        4 &
        $Z\rightarrow e^+e^-$,
        $J/\psi\rightarrow e^+e^-$ &
        Electron tracking &
        $\varepsilon_e$ (Eq.~\ref{eq:electron_eff}),
        momentum smearing (Eq.~\ref{eq:mom_smear})
        \\
        \bottomrule
    \end{tabular}
    \caption{
        Sequential tuning procedure for the detector-response parameters.
        At each step, only the parameters of the indicated block are trainable;
        previously tuned blocks are fixed, and all other parameters remain at
        their default (i.e. initial) CMS card values.
    }
    \label{tab:tuning_procedure}
\end{table}

% First, the $Z\rightarrow\mu^+\mu^-$ and $J/\psi\rightarrow\mu^+\mu^-$ samples are used to tune the muon tracking efficiency parameters of Eq.\ref{eq:muon_eff} and the muon momentum smearing parameters of Eq.~\ref{eq:mom_smear}. The $K^0_S \rightarrow \pi^+ \pi^-$ sample is then used to tune the charged hadron tracking parameters of Eq.~\ref{eq:chad_eff} and \ref{eq:mom_smear}, the dijet sample to tune the ECal and HCal parameters of Eq.~\ref{eq:ecal_res} and \ref{eq:hcal_res}, and finally the $Z\rightarrow e^+e^-$ and $J/\psi\rightarrow e^+e^-$ samples to tune the electron tracking parameters of Eq.~\ref{eq:electron_eff} and \ref{eq:mom_smear}. 
% At each step only the parameters of the block being tuned are trainable. 
% The blocks tuned in earlier steps are fixed at their tuned values and all remaining parameters are held at the default CMS card values.

Sequential tuning is possible because the parameter blocks act on disjoint sets of objects, and the samples are chosen to isolate them. Each charged particle species has its own unique set of tracking and momentum smearing parameters, and the calorimeters act on every particle but muons. A parameter that acts on no object in a sample leaves that sample's distributions and loss unchanged. For samples that interact with several blocks, the stages are ordered so that all or most parameters other than the subset being tuned are already fixed at their fitted values. For example, the muon tracking and smearing parameters are tuned first because muons do not interact with the calorimeters. Then, the charged hadron tracking and smearing parameters are fit before the calorimeter parameters, since the calorimeters use the charged hadron momenta and energies. Likewise, the calorimeter parameters are fit before the tracking and smearing parameters for electrons, whose reconstructed energy combines track and ECal information; although the calorimeters do interact with electrons, the QCD dijet sample provides sufficient signal from other particles (i.e. charged hadrons, neutral hadrons, and photons) to tune the calorimeter parameters. In this way, each block is determined by one stage and carried forward without iteration.

\begin{figure}[htbp]
  \centering
  \captionsetup[sub]{justification=centering, singlelinecheck=false}

  \begin{subfigure}[t]{0.24\textwidth}
    \vspace{0pt}\centering
    \includegraphics[height=8cm,page=2]{images/params_summary.pdf}
    \caption{Charged Hadron Parameters}\label{fig:params_reg_a}
  \end{subfigure}\hfill
  \begin{subfigure}[t]{0.24\textwidth}
    \vspace{0pt}\centering
    \includegraphics[height=8cm,page=1]{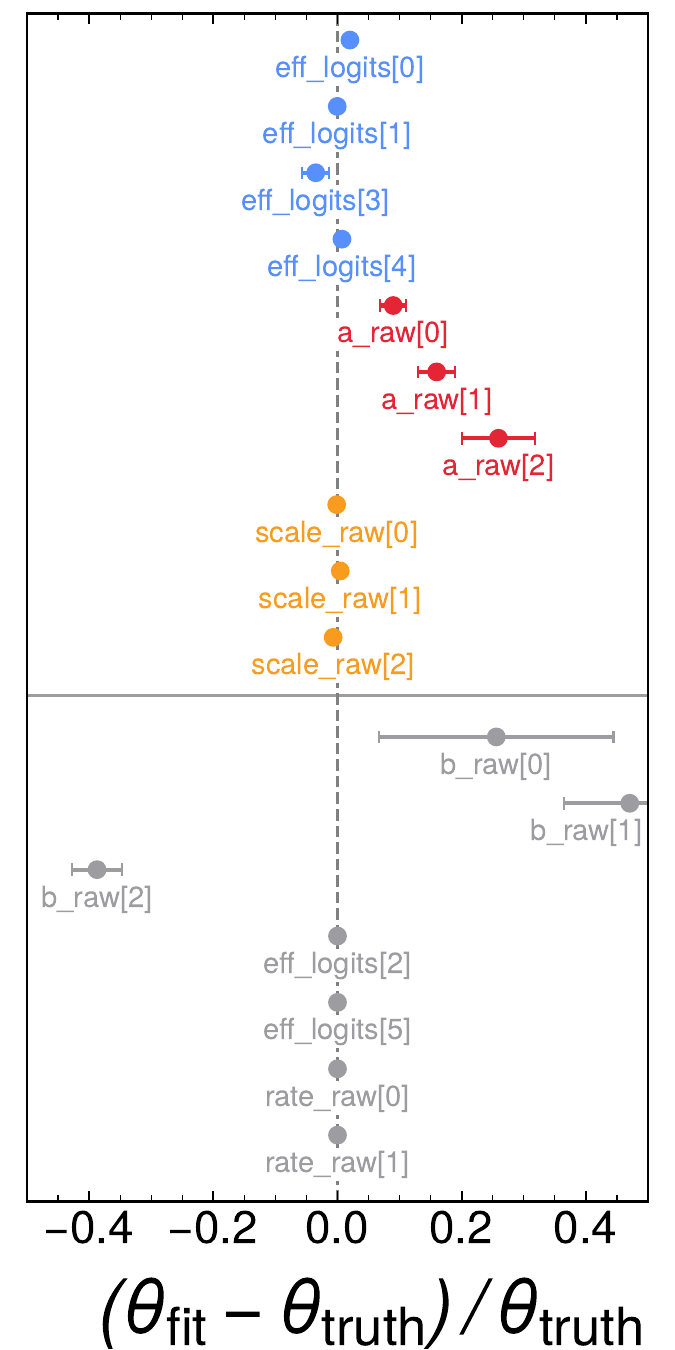}
    \caption{Muon Parameters}\label{fig:params_reg_b}
  \end{subfigure}\hfill
  \begin{subfigure}[t]{0.24\textwidth}
    \vspace{0pt}\centering
    \includegraphics[height=8cm,page=3]{images/params_summary.pdf}
    \caption{Electron Parameters}\label{fig:params_reg_c}
  \end{subfigure}\hfill
  \begin{subfigure}[t]{0.24\textwidth}
    \vspace{0pt}\centering
    \includegraphics[height=8cm,page=4]{images/params_summary.pdf}
    \caption{Calorimeter Parameters}\label{fig:params_reg_d}
  \end{subfigure}

  \caption{Parameter regression result for different modules of the delphes card with the mean represented as a marker and the standard deviation represented as an error bar. Fig~\ref{fig:params_reg_a} shows the charged hadron parameters, which are tuned using $K^0_s\rightarrow\pi^+\pi^-$ samples. Fig~\ref{fig:params_reg_b} shows the muon parameters, which are tuned using $Z\rightarrow\mu^+\mu^-$ and $J/\psi\rightarrow\mu^+\mu^-$ samples. Fig~\ref{fig:params_reg_c} shows the electron parameters, which are tuned using $Z\rightarrow e^+e^-$ and $J/\psi\rightarrow e^+e^-$ samples. Fig~\ref{fig:params_reg_d} shows the calorimeter parameters, which are tuned using dijet samples. Most of the 62 trainable parameters are close to the true value, with the exceptions due to low statistics in high $p_\mathrm{T}$ or forward regions colored in grey.}
  \label{fig:params_reg}
\end{figure}

The result of the parameter regression is shown in Fig.~\ref{fig:params_reg}. 
Five Parnassus-P models are tuned with different random seeds in training, and the mean and standard deviation of fitted parameters are taken.
Most of the perturbed parameters are recovered to their true values with a few exceptions. 
These are parameters that act mainly at high $p_\mathrm{T}$ or in the forward region, where the training samples contain few events, so their fitted values deviate substantially from the truth. 
For example, the $\mathtt{b\_raw}$ parameters of Eq.~\ref{eq:mom_smear} enter the relative momentum resolution as $(b ~ p_\mathrm{T})^2$, which are sub-dominant to the constant term $\mathtt{a\_raw}$ when acting on leptons from $J/\psi$ and $Z$ decays and pions from relatively low $p_\mathrm{T}$ $K^0_S$ decays. 
The loss is only weakly sensitive and its gradient is correspondingly small.

\begin{figure}[htbp]
  \centering
  % ---- row 1 ----
  \begin{subfigure}[b]{0.32\textwidth}
    \includegraphics[width=\linewidth,page=1]{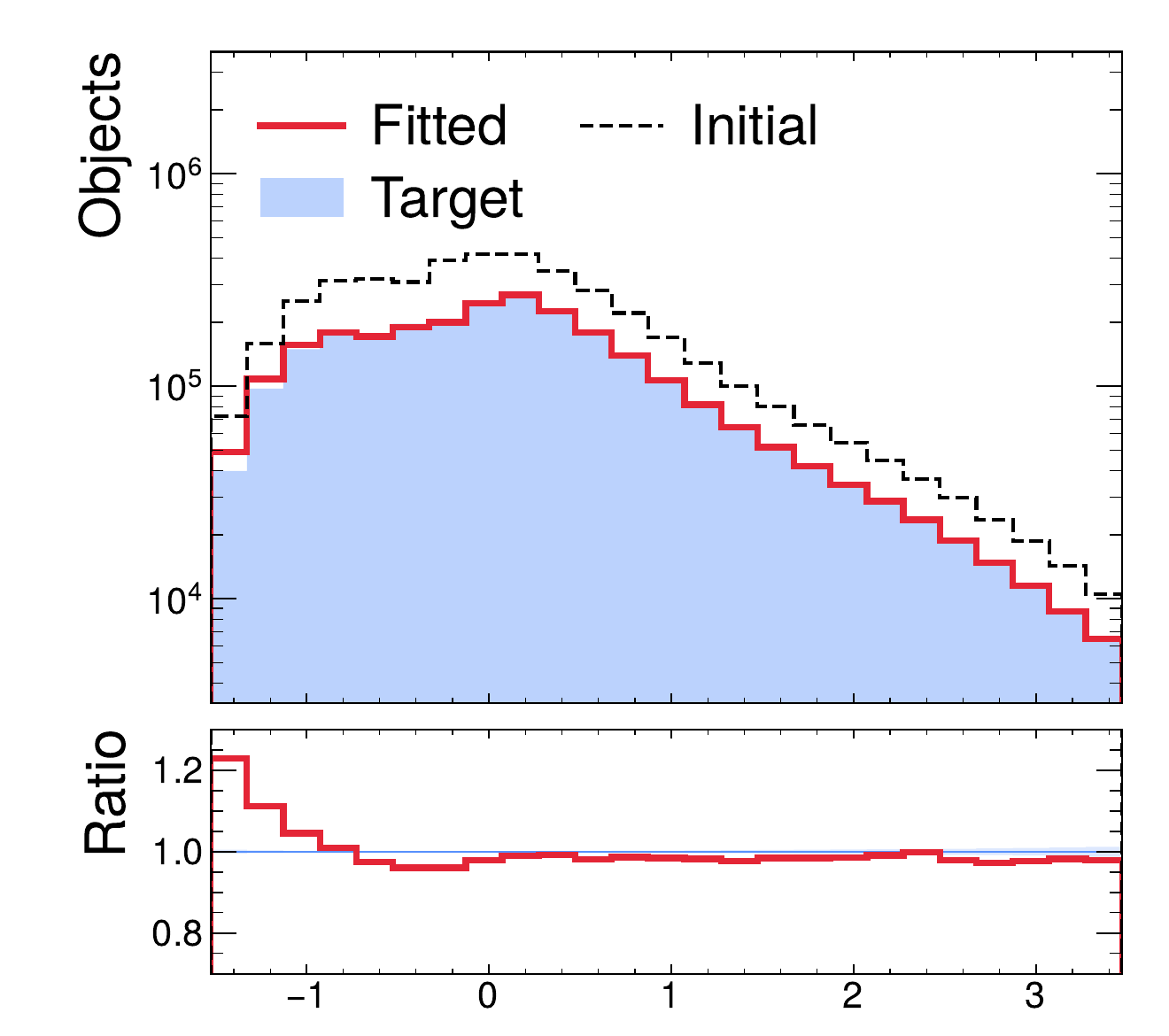}
    \caption{Charged Hadron\\log($p_\mathrm{T}$/GeV)}\label{fig:HZZ4l_a}
  \end{subfigure}\hfill
  \begin{subfigure}[b]{0.32\textwidth}
    \includegraphics[width=\linewidth,page=4]{images/HZZ4l_pseudodata.pdf}
    \caption{Neutral Hadron\\log($p_\mathrm{T}$/GeV)}\label{fig:HZZ4l_b}
  \end{subfigure}\hfill
  \begin{subfigure}[b]{0.32\textwidth}
    \includegraphics[width=\linewidth,page=7]{images/HZZ4l_pseudodata.pdf}
    \caption{Electron\\log($p_\mathrm{T}$/GeV)}\label{fig:HZZ4l_c}
  \end{subfigure}

  \vspace{1em}
  % ---- row 2 ----
  \begin{subfigure}[b]{0.32\textwidth}
    \includegraphics[width=\linewidth,page=11]{images/HZZ4l_pseudodata.pdf}
    \caption{Muon\\log($p_\mathrm{T}$/GeV)}\label{fig:HZZ4l_d}
  \end{subfigure}\hfill
  \begin{subfigure}[b]{0.32\textwidth}
    \includegraphics[width=\linewidth,page=14]{images/HZZ4l_pseudodata.pdf}
    \caption{Photon\\log($p_\mathrm{T}$/GeV)}\label{fig:HZZ4l_e}
  \end{subfigure}\hfill
    \begin{subfigure}[b]{0.32\textwidth}
    \includegraphics[width=\linewidth,page=21]{images/HZZ4l_pseudodata.pdf}
    \caption{$m_{4l}$/GeV}\label{fig:HZZ4l_f}
  \end{subfigure}\hfill

  \caption{Kinematic feature distributions of $H\rightarrow ZZ^{(\ast)} \rightarrow 4l$ events. The initial distributions are produced by the default Delphes CMS card. The target distributions are produced by a card that is perturbed from this default CMS configuration. The fitted distributions are produced by the Parnassus-P model that was initialized at the default CMS configuration, and trained on the data generated by the perturbed/target configuration. Fig.~\ref{fig:HZZ4l_a} to \ref{fig:HZZ4l_e} show the $p_\mathrm{T}$ distribution of pflow objects, and Fig.~\ref{fig:HZZ4l_f} shows the distribution of resonant mass of four leading leptons.}
  \label{fig:HZZ4l_distribution}
\end{figure}

The accuracy of the regressed parameters is further validated on a held-out $H \rightarrow ZZ^{(\ast)} \rightarrow 4\ell$ sample. 
The sequentially tuned Parnassus-P card, which never saw this sample, is used to simulate the detector response for 100,000 $H \rightarrow ZZ^{(\ast)} \rightarrow 4\ell$ events. 
The distributions of the reconstructed objects and high-level observables are compared with those obtained from the default CMS card and from the perturbed card used to generate the pseudo-data, which serves as the target. 
The results are shown in Fig.~\ref{fig:HZZ4l_distribution}. 
For all observables the tuned card reproduces the target distributions closely, whereas the default card shows visible deviations.

\subsection{Towards a Fit to CMS Data}
\label{sec:opendata}

\begin{figure}[htbp]
  \centering
  \begin{subfigure}[b]{0.44\textwidth}
    \includegraphics[width=\linewidth,page=1]{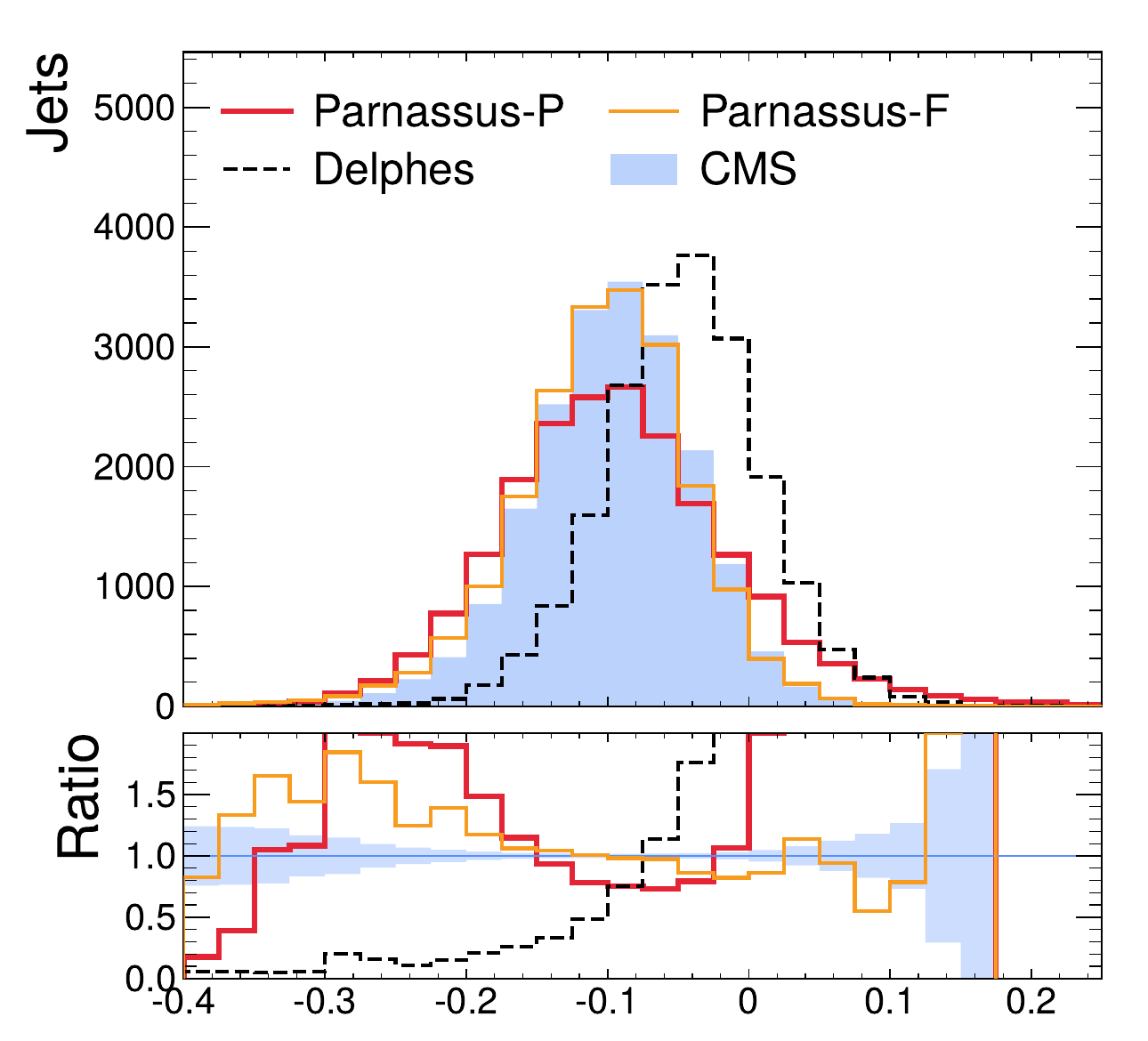}
    \caption{$(p^\mathrm{reco}_\mathrm{T ~ jet} - p^\mathrm{truth}_\mathrm{T ~ jet}) / p^\mathrm{truth}_\mathrm{T ~ jet}$}\label{fig:fullsim_comp_g}
  \end{subfigure}\hfill
  \begin{subfigure}[b]{0.44\textwidth}
    \includegraphics[width=\linewidth,page=2]{images/fullsim_comparison_jetlevel.pdf}
    \caption{$(m^\mathrm{reco}_\mathrm{jet} - m^\mathrm{truth}_\mathrm{jet}) / m^\mathrm{truth}_\mathrm{jet}$}\label{fig:fullsim_comp_h}
  \end{subfigure}\hfill
  \begin{subfigure}[b]{0.44\textwidth}
    \includegraphics[width=\linewidth,page=3]{images/fullsim_comparison_jetlevel.pdf}
    \caption{$\eta^\mathrm{reco}$ - $\eta^\mathrm{truth}$}\label{fig:fullsim_comp_i}
  \end{subfigure}\hfill
  \begin{subfigure}[b]{0.44\textwidth}
    \includegraphics[width=\linewidth,page=4]{images/fullsim_comparison_jetlevel.pdf}
    \caption{$\phi^\mathrm{reco}$ - $\phi^\mathrm{truth}$}\label{fig:fullsim_comp_j}
  \end{subfigure}\hfill
  
  \caption{Kinematic distribution of reconstructed leading jets simulated by Parnassus-P, Parnassus-F, Delphes, and CMS full simulation.}
  % Fig. \ref{fig:fullsim_comp_a} - \ref{fig:fullsim_comp_c} show the kinematic distribution of particle-flow objects and the event multiplicity.
  \label{fig:fullsim_comparison}
\end{figure}

Parnassus-P can also be used to tune Delphes to a full detector simulation, in which case no true parameter configuration exists. 
To demonstrate this, the Parnassus-P model is trained on CMS open simulation data of light jets~\cite{komiskeExploringSpaceJets2020,dreyer_2024_11389651}, consisting of particle-flow candidates of one anti-$k_\mathrm{T}$ $R = 0.5$~\cite{Cacciari:2008gp} jet together with the generator-level particles in the same cone. 
The generator-level particles form the input to Parnassus-P and the particle-flow candidates form the target. 
The model is trained on 100,000 jets from the $\hat{p}_\mathrm{T} > 1$~TeV sample and validated on a held-out set of 20,000 jets from the same sample; the validation results are shown in Fig.~\ref{fig:fullsim_comparison}.

To mitigate some of the modeling discrepancies between Delphes and CMS full simulation, a "full-sim" version of Parnassus-P is selectable through a user argument. This version makes three main adjustments to the model described in Section~\ref{sec:parametrization}. First, the charged-hadron tracking efficiency of Eq.~\ref{eq:chad_eff} is given a finer $p_\mathrm{T}$ binning with edges at $(0.1, 1, 10, 25, 50, 100)$~GeV. 
This gives the model the flexibility to describe the $p_\mathrm{T}$ dependence of the charged hadron yield observed in full simulation. 
Second, a photon merger module is added that clusters reconstructed photons within $\Delta R < R$ into a single object, mimicking the clustering of electromagnetic deposits in the particle-flow reconstruction. 
$R = 0.09$ is chosen by a scan to minimise the validation loss. 
Third, the full simulation contains reconstructed charged hadrons with no counterpart in the generator record, such as secondary particles from interactions with the detector material. These effects are not reproducible by Parnassus-P, so a $p_\mathrm{T}$ threshold of 5~GeV and $|\eta| < 2.7$ is applied on all reconstructed objects to remove the soft part of the full simulation.
All 62 parameters of the Parnassus-P card are tuned simultaneously.

The performance of Parnassus-P model is compared with that of the pretrained Parnassus-F model and of Delphes with the CMS card, all applied to jets from the same physics process.
The Delphes reconstruction with the default CMS card includes pileup with an average of 6.35 pileup vertices per event, and Parnassus-F model is trained on samples with pileup. The Parnassus-P model does not include pileup in simulation\footnote{This is not a fundamental constraint and future studies will mitigate the impact of pileup.}; however, since only particle-flow candidates inside the jet cone with $p_\mathrm{T} > 5$~GeV are retained, the residual effect of pileup on the comparison is expected to be negligible.

For each model, 100,000 dijet events are processed and the throughput is recorded. Parnassus-P and Parnassus-F are run on a single NVIDIA A100 GPU with a batch size of 2048, reaching 22,808 and 135 events per second, respectively, while Delphes is run on a single AMD EPYC 7713 CPU core and processes 118 events per second.
Compared to Delphes and Parnassus-F, Parnassus-P benefits from the same simplified detector-simulation operations as Delphes, but vectorizes over all particles of a batch on the GPU.
As a result, the throughput of Parnassus-P exceeds the throughput of Parnassus-F and Delphes by more than two orders of magnitude\footnote{This will slightly change with pileup, but since the pileup is overlaid, it is mostly a fixed cost.}.

The responses of the reconstructed jet variables are shown in Fig.~\ref{fig:fullsim_comparison}.
The tuned Parnassus-P model shows better agreement with respect to full simulation than the default Delphes card, but not as good as Parnassus-F. 
The remaining gap reflects effects present in the full simulation that have no counterpart in the parametric model of Delphes, and therefore cannot be absorbed by any setting of its 62 tunable parameters, whereas Parnassus-F has no such functional limitations.

\section{Conclusion and Outlook}
\label{sec:conclusions}

In this paper, we have promoted the parametric model within Parnassus to be differentiable, performed closure tests against pseudodata to validate its performance and differentiability, and performed the first fits against CMS full simulation.  The resulting Parnassus card is more accurate than the default Delphes card by construction and inherits all the benefits of the Parnassus framework, including being Python-based and GPU native.  The parametric path complements the more accurate, but less interpretable and slower neural-network path.  

A number of improvements are envisioned for Parnassus-P in the future.  From our experience with the closure test, we learned that pinning down the exact parameters requires a diverse set of carefully chosen datasets.  Since the training uses simulations, we can engineer datasets for more complete phase-space coverage and better performance in general.  For example, a simulated, high-mass $Z'$ may help constrain the high-$p_T$ parameters. Modifying the parametric framework to add more parameters (including replacing formulas with neural networks) and reduce mismodeling effects with respect to full simulation is also straightforward and could enhance accuracy.

All of our software and datasets are publicly available, and we welcome suggestions for new features via Github issues, as well as new Parnassus-P or Parnassus-F models submitted via merge requests.

\section*{Data and Code Availability}

%The code for this paper can be found at \url{https://github.com/parnassus-hep/cms-flow-evt}. 
The postprocessed CMS dataset and the generated \textsc{Delphes} dataset can be found on Zenodo at \url{https://zenodo.org/records/15083495}. The closure test dataset can be found on Zenodo at \url{https://doi.org/10.5281/zenodo.22071385}. The code is available at \url{https://github.com/abdelabd/parnassus/tree/diff_delphes_runze_cmssinglejet}.

\section*{Acknowledgments}
AE is supported by a DOE CompHEP Western Advanced Training for Computational High-Energy Physics (WATCHEP) Fellowship, under DOE Award DE-SC-0023527. RL is supported by an MCGen Fellowship, under NSF Award OAC-2417682.  BN is supported by the U.S. Department of Energy (DOE) under Contract No. DE-AC02-76SF00515 and by the CompHEP Center for Computational Excellence. DK and EG are supported by the Minerva Grant, the Knell Family Institute for Artificial Intelligence, a BSF Grant, and the Krenter-Perinot Center for High Energy Particle Physics.   This research used resources of the National Energy Research Scientific Computing Center (NERSC), a DOE Office of Science User Facility supported by the Office of Science of the U.S. Department of Energy under Contract No. DE-AC02-05CH11231, under NERSC award HEP-ERCAP0035546.  This research also used the Delta advanced computing and data resource which is supported by the National Science Foundation (award OAC 2005572) and the State of Illinois. Delta is a joint effort of the University of Illinois Urbana-Champaign and its National Center for Supercomputing Applications.

We acknowledge the use of Claude Code (Anthropic) throughout this project—for brainstorming and refining the central idea, for collaborating on the theoretical results, for assistance in scripting and running the numerical studies, and for help editing the manuscript—and we take full responsibility for the content of this manuscript.

\bibliography{references}

@book{villani2009optimal,
  title     = {Optimal Transport: Old and New},
  author    = {Villani, C{\'e}dric},
  series    = {Grundlehren der mathematischen Wissenschaften},
  volume    = {338},
  publisher = {Springer},
  year      = {2009},
  doi       = {10.1007/978-3-540-71050-9}
}

@misc{akiba2019optuna,
      title={Optuna: A Next-generation Hyperparameter Optimization Framework}, 
      author={Takuya Akiba and Shotaro Sano and Toshihiko Yanase and Takeru Ohta and Masanori Koyama},
      year={2019},
      eprint={1907.10902},
      archivePrefix={arXiv},
      primaryClass={cs.LG},
      url={https://arxiv.org/abs/1907.10902}, 
}

@misc{adam2017,
      title={Adam: A Method for Stochastic Optimization}, 
      author={Diederik P. Kingma and Jimmy Ba},
      year={2017},
      eprint={1412.6980},
      archivePrefix={arXiv},
      primaryClass={cs.LG},
      url={https://arxiv.org/abs/1412.6980}, 
}

@article{Cacciari:2008gp,
    author = "Cacciari, Matteo and Salam, Gavin P. and Soyez, Gregory",
    title = "{The anti-$k_t$ jet clustering algorithm}",
    eprint = "0802.1189",
    archivePrefix = "arXiv",
    primaryClass = "hep-ph",
    reportNumber = "LPTHE-07-03",
    doi = "10.1088/1126-6708/2008/04/063",
    journal = "JHEP",
    volume = "04",
    pages = "063",
    year = "2008"
}

@article{Krause:2026ayh,
    author = "Krause, Claudius and Winterhalder, Ramon and Feickert, Matthew and Nachman, Benjamin",
    title = "{The Living Guide of Machine Learning for Particle Physics}",
    eprint = "2608.09531",
    archivePrefix = "arXiv",
    primaryClass = "hep-ph",
    reportNumber = "MBI-ML-26-05, TIF-UNIMI-2026-10",
    month = "8",
    year = "2026"
}

@article{Buckley:2009bj,
    author = "Buckley, Andy and Hoeth, Hendrik and Lacker, Heiko and Schulz, Holger and von Seggern, Jan Eike",
    title = "{Systematic event generator tuning for the LHC}",
    eprint = "0907.2973",
    archivePrefix = "arXiv",
    primaryClass = "hep-ph",
    reportNumber = "IPPP-09-52, DCPT-104-22, LU-TP-09-18, HU-EP-09-33, MCNET-09-14",
    doi = "10.1140/epjc/s10052-009-1196-7",
    journal = "Eur. Phys. J. C",
    volume = "65",
    pages = "331--357",
    year = "2010"
}

@misc{aehle2024efficientforwardmodealgorithmicderivatives,
      title={Efficient Forward-Mode Algorithmic Derivatives of Geant4}, 
      author={Max Aehle and Xuan Tung Nguyen and Mihály Novák and Tommaso Dorigo and Nicolas R. Gauger and Jan Kieseler and Markus Klute and Vassil Vassilev},
      year={2024},
      eprint={2407.02966},
      archivePrefix={arXiv},
      primaryClass={physics.comp-ph},
      url={https://arxiv.org/abs/2407.02966}, 
}

@misc{aehle2024optimizationusingpathwisealgorithmic,
      title={Optimization Using Pathwise Algorithmic Derivatives of Electromagnetic Shower Simulations}, 
      author={Max Aehle and Mihály Novák and Vassil Vassilev and Nicolas R. Gauger and Lukas Heinrich and Michael Kagan and David Lange},
      year={2024},
      eprint={2405.07944},
      archivePrefix={arXiv},
      primaryClass={physics.comp-ph},
      url={https://arxiv.org/abs/2405.07944}, 
}

@article{Alterkait:2026ocv,
    author = "Alterkait, Omar and Jes{\'u}s-Valls, C{\'e}sar and Matsumoto, Ryo and de Perio, Patrick and Terao, Kazuhiro",
    title = "{End-to-end Differentiable Calibration and Reconstruction for Optical Particle Detectors}",
    eprint = "2602.24129",
    archivePrefix = "arXiv",
    primaryClass = "hep-ex",
    month = "2",
    year = "2026"
}

@article{Gasiorowski:2023tqf,
    author = "Gasiorowski, Sean and Chen, Yifan and Nashed, Youssef and Granger, Pierre and Mironov, Camelia and Tsang, Ka Vang and Ratner, Daniel and Terao, Kazuhiro",
    title = "{Differentiable simulation of a liquid argon time projection chamber}",
    eprint = "2309.04639",
    archivePrefix = "arXiv",
    primaryClass = "physics.ins-det",
    doi = "10.1088/2632-2153/ad2cf0",
    journal = "Mach. Learn. Sci. Tech.",
    volume = "5",
    number = "2",
    pages = "025012",
    year = "2024"
}

@article{MODE:2023ekf,
    author = "Aehle, Max and others",
    collaboration = "MODE",
    title = "{Progress in end-to-end optimization of fundamental physics experimental apparata with differentiable programming}",
    eprint = "2310.05673",
    archivePrefix = "arXiv",
    primaryClass = "physics.ins-det",
    reportNumber = "FERMILAB-PUB-23-608-CSAID-PPD",
    doi = "10.1016/j.revip.2025.100120",
    journal = "Rev. Phys.",
    volume = "13",
    pages = "100120",
    year = "2025"
}

@article{MODE:2022znx,
    author = "Dorigo, Tommaso and others",
    collaboration = "MODE",
    title = "{Toward the end-to-end optimization of particle physics instruments with differentiable programming}",
    eprint = "2203.13818",
    archivePrefix = "arXiv",
    primaryClass = "physics.ins-det",
    doi = "10.1016/j.revip.2023.100085",
    journal = "Rev. Phys.",
    volume = "10",
    pages = "100085",
    year = "2023"
}

@inproceedings{Adelmann:2022ozp,
    author = "Adelmann, Andreas and others",
    title = "{New directions for surrogate models and differentiable programming for High Energy Physics detector simulation}",
    booktitle = "{Snowmass 2021}",
    eprint = "2203.08806",
    archivePrefix = "arXiv",
    primaryClass = "hep-ph",
    reportNumber = "FERMILAB-CONF-22-199-SCD",
    month = "3",
    year = "2022"
}

@article{Feickert:2021ajf,
    author = "Feickert, Matthew and Nachman, Benjamin",
    title = "{A Living Review of Machine Learning for Particle Physics}",
    eprint = "2102.02770",
    archivePrefix = "arXiv",
    primaryClass = "hep-ph",
    month = "2",
    year = "2021"
}

@article{Elabd:2026qfw,
    author = "Elabd, Abdelrahman and Gross, Eilam and Kobylianskii, Dmitrii and Nachman, Benjamin",
    title = "{Parnassus: A GPU-enabled, Python-based Package for Fast Particle Detector Simulation and Reconstruction}",
    eprint = "2606.26089",
    archivePrefix = "arXiv",
    primaryClass = "hep-ex",
    month = "6",
    year = "2026"
}

@article{Butter:2022rso,
    author = "Badger, Simon and others",
    editor = "Butter, Anja and Plehn, Tilman and Schumann, Steffen",
    title = "{Machine learning and LHC event generation}",
    eprint = "2203.07460",
    archivePrefix = "arXiv",
    primaryClass = "hep-ph",
    reportNumber = "FERMILAB-PUB-22-126-T",
    doi = "10.21468/SciPostPhys.14.4.079",
    journal = "SciPost Phys.",
    volume = "14",
    number = "4",
    pages = "079",
    year = "2023"
}

@article{Hashemi:2023rgo,
    author = "Hashemi, Baran and Krause, Claudius",
    title = "{Deep generative models for detector signature simulation: A taxonomic review}",
    eprint = "2312.09597",
    archivePrefix = "arXiv",
    primaryClass = "physics.ins-det",
    reportNumber = "HEPHY-ML-23-02",
    doi = "10.1016/j.revip.2024.100092",
    journal = "Rev. Phys.",
    volume = "12",
    pages = "100092",
    year = "2024"
}

@article{Krause:2024avx,
    author = "Amram, Oz and others",
    editor = "Krause, Claudius and Faucci Giannelli, Michele and Kasieczka, Gregor and Nachman, Benjamin and Salamani, Dalila and Shih, David and Zaborowska, Anna",
    title = "{CaloChallenge 2022: a community challenge for fast calorimeter simulation}",
    eprint = "2410.21611",
    archivePrefix = "arXiv",
    primaryClass = "physics.ins-det",
    reportNumber = "HEPHY-ML-24-05, FERMILAB-PUB-24-0728-CMS, TTK-24-43",
    doi = "10.1088/1361-6633/ae1304",
    journal = "Rept. Prog. Phys.",
    volume = "88",
    number = "11",
    pages = "116201",
    year = "2025"
}

@article{Paganini:2017dwg,
    author = "Paganini, Michela and de Oliveira, Luke and Nachman, Benjamin",
    title = "{CaloGAN : Simulating 3D high energy particle showers in multilayer electromagnetic calorimeters with generative adversarial networks}",
    eprint = "1712.10321",
    archivePrefix = "arXiv",
    primaryClass = "hep-ex",
    doi = "10.1103/PhysRevD.97.014021",
    journal = "Phys. Rev. D",
    volume = "97",
    number = "1",
    pages = "014021",
    year = "2018"
}

@article{Paganini:2017hrr,
    author = "Paganini, Michela and de Oliveira, Luke and Nachman, Benjamin",
    title = "{Accelerating Science with Generative Adversarial Networks: An Application to 3D Particle Showers in Multilayer Calorimeters}",
    eprint = "1705.02355",
    archivePrefix = "arXiv",
    primaryClass = "hep-ex",
    doi = "10.1103/PhysRevLett.120.042003",
    journal = "Phys. Rev. Lett.",
    volume = "120",
    number = "4",
    pages = "042003",
    year = "2018"
}

@misc{lipman2023flowmatchinggenerativemodeling,
      title={Flow Matching for Generative Modeling}, 
      author={Yaron Lipman and Ricky T. Q. Chen and Heli Ben-Hamu and Maximilian Nickel and Matt Le},
      year={2023},
      eprint={2210.02747},
      archivePrefix={arXiv},
      primaryClass={cs.LG},
      url={https://arxiv.org/abs/2210.02747}, 
}

@article{Lo:2026use,
    author = "Lo, Ya-Feng and Kobylianskii, Dmitrii and Nachman, Benjamin and Gross, Eilam",
    title = "{An AI-based Detector Simulation and Reconstruction Model for the ALEPH Experiment at LEP}",
    eprint = "2604.11834",
    archivePrefix = "arXiv",
    primaryClass = "physics.ins-det",
    month = "4",
    year = "2026"
}

@article{Dreyer:2024bhs,
    author = "Dreyer, Etienne and Gross, Eilam and Kobylianskii, Dmitrii and Mikuni, Vinicius and Nachman, Benjamin and Soybelman, Nathalie",
    title = "{Automated Approach to Accurate, Precise, and Fast Detector Simulation and Reconstruction}",
    eprint = "2406.01620",
    archivePrefix = "arXiv",
    primaryClass = "physics.data-an",
    doi = "10.1103/PhysRevLett.133.211902",
    journal = "Phys. Rev. Lett.",
    volume = "133",
    number = "21",
    pages = "211902",
    year = "2024"
}

@article{Kobylianskii:2024sup,
    author = "Kobylianskii, Dmitrii and Soybelman, Nathalie and Kakati, Nilotpal and Dreyer, Etienne and Nachman, Benjamin and Gross, Eilam",
    title = "{Advancing set-conditional set generation: Diffusion models for fast simulation of reconstructed particles}",
    eprint = "2405.10106",
    archivePrefix = "arXiv",
    primaryClass = "hep-ex",
    doi = "10.1103/PhysRevD.110.092013",
    journal = "Phys. Rev. D",
    volume = "110",
    number = "9",
    pages = "092013",
    year = "2024"
}

@article{Dreyer:2025zhp,
    author = "Dreyer, Etienne and Gross, Eilam and Kobylianskii, Dmitrii and Mikuni, Vinicius and Nachman, Benjamin",
    title = "{Conditional deep generative models for simultaneous simulation and reconstruction of entire events}",
    eprint = "2503.19981",
    archivePrefix = "arXiv",
    primaryClass = "hep-ex",
    doi = "10.1103/14ph-482n",
    journal = "Phys. Rev. D",
    volume = "113",
    number = "3",
    pages = "032005",
    year = "2026"
}

@article{GEANT4:2002zbu,
    author = "Agostinelli, S. and others",
    collaboration = "GEANT4",
    title = "{GEANT4 - A Simulation Toolkit}",
    reportNumber = "SLAC-PUB-9350, FERMILAB-PUB-03-339, CERN-IT-2002-003",
    doi = "10.1016/S0168-9002(03)01368-8",
    journal = "Nucl. Instrum. Meth. A",
    volume = "506",
    pages = "250--303",
    year = "2003"
}

@article{deFavereau:2013fsa,
    author = "de Favereau, J. and Delaere, C. and Demin, P. and Giammanco, A. and Lema{\^\i}tre, V. and Mertens, A. and Selvaggi, M.",
    collaboration = "DELPHES 3",
    title = "{DELPHES 3, A modular framework for fast simulation of a generic collider experiment}",
    eprint = "1307.6346",
    archivePrefix = "arXiv",
    primaryClass = "hep-ex",
    doi = "10.1007/JHEP02(2014)057",
    journal = "JHEP",
    volume = "02",
    pages = "057",
    year = "2014"
}

@dataset{dreyer_2024_11389651,
  author       = {Dreyer, Etienne and
                  Gross, Eilam and
                  Kobylianskii, Dmitrii and
                  Mikuni, Vinicius and
                  Nachman, Benjamin and
                  Soybelman, Nathalie},
  title        = {Simulated and Reconstructed Jets with CMS and
                   DELPHES
                  },
  month        = may,
  year         = 2024,
  publisher    = {Zenodo},
  doi          = {10.5281/zenodo.11389651},
  url          = {https://doi.org/10.5281/zenodo.11389651},
}

@article{sjostrandBriefIntroductionPYTHIA2008,
    title = {A brief introduction to {PYTHIA} 8.1},
    volume = {178},
    issn = {0010-4655},
    url = {https://www.sciencedirect.com/science/article/pii/S0010465508000441},
    doi = {10.1016/j.cpc.2008.01.036},
    number = {11},
    urldate = {2024-04-14},
    journal = {Computer Physics Communications},
    author = {Sjöstrand, Torbjörn and Mrenna, Stephen and Skands, Peter},
    month = jun,
    year = {2008},
    pages = {852--867},
}

@article{bierlichComprehensiveGuidePhysics2022,
    title = {A comprehensive guide to the physics and usage of {PYTHIA} 8.3},
    issn = {2949-804X},
    url = {https://scipost.org/SciPostPhysCodeb.8},
    doi = {10.21468/SciPostPhysCodeb.8},
    language = {en},
    urldate = {2026-06-02},
    journal = {SciPost Physics Codebases},
    author = {Bierlich, Christian and Chakraborty, Smita and Desai, Nishita and Gellersen, Leif and Helenius, Ilkka and Ilten, Philip and Lönnblad, Leif and Mrenna, Stephen and Prestel, Stefan and Preuss, Christian Tobias and Sjöstrand, Torbjörn and Skands, Peter and Utheim, Marius and Verheyen, Rob},
    month = nov,
    year = {2022},
    pages = {008},
}

@article{CMS:2017yfk,
    author = "Sirunyan, A. M. and others",
    collaboration = "CMS",
    title = "{Particle-flow reconstruction and global event description with the CMS detector}",
    eprint = "1706.04965",
    archivePrefix = "arXiv",
    primaryClass = "physics.ins-det",
    reportNumber = "CMS-PRF-14-001, CERN-EP-2017-110",
    doi = "10.1088/1748-0221/12/10/P10003",
    journal = "JINST",
    volume = "12",
    number = "10",
    pages = "P10003",
    year = "2017"
}

@article{ATLAS:2017ghe,
    author = "Aaboud, Morad and others",
    collaboration = "ATLAS",
    title = "{Jet reconstruction and performance using particle flow with the ATLAS Detector}",
    eprint = "1703.10485",
    archivePrefix = "arXiv",
    primaryClass = "hep-ex",
    reportNumber = "CERN-EP-2017-024",
    doi = "10.1140/epjc/s10052-017-5031-2",
    journal = "Eur. Phys. J. C",
    volume = "77",
    number = "7",
    pages = "466",
    year = "2017"
}

@article{komiskeExploringSpaceJets2020,
    title = {Exploring the {Space} of {Jets} with {CMS} {Open} {Data}},
    volume = {101},
    issn = {2470-0010, 2470-0029},
    url = {http://arxiv.org/abs/1908.08542},
    doi = {10.1103/PhysRevD.101.034009},
    number = {3},
    urldate = {2024-02-26},
    journal = {Physical Review D},
    author = {Komiske, Patrick T. and Mastandrea, Radha and Metodiev, Eric M. and Naik, Preksha and Thaler, Jesse},
    month = feb,
    year = {2020},
    note = {arXiv:1908.08542 [hep-ex, physics:hep-ph]},
    pages = {034009},
}
\bibliographystyle{apsrev4-1}

\clearpage
\onecolumngrid

\appendix

\end{document}